\documentclass[aps,showpacs,prl,twocolumn]{revtex4-1}

\usepackage{graphicx}
\usepackage{amsmath, amsthm, amssymb}
\usepackage[caption=false]{subfig} 
\usepackage{color}

\newcommand*{\myT}{{T}}

\newcommand{\Lfun}[2]{{\cal L}^{(#1)}_{#2}}

\begin{document}

\title{Tunable Quantum Temperature Oscillations in Graphene and Carbon Nanoribbons}

\author{Justin P. Bergfield}
\author{Mark A. Ratner}
\affiliation{Department of Chemistry, Northwestern University, 2145 Sheridan Road, Evanston, IL, 60208}

\author{Charles A. Stafford}
\affiliation{Department of Physics, University of Arizona, 1118 East Fourth Street, Tucson, AZ 85721}

\author{Massimiliano Di Ventra}
\affiliation{Department of Physics, University of California, San Diego, La Jolla, CA 92093}

\date{\today}
\begin{abstract}
We investigate the local electron temperature distribution
in carbon nanoribbon (CNR) and graphene junctions subject to
an applied thermal gradient.  Using a realistic model of a scanning thermal microscope, we predict
quantum temperature oscillations whose
wavelength is related to that of Friedel oscillations.  Experimentally,
this wavelength can be tuned over several orders of magnitude by gating/doping, bringing quantum temperature oscillations
within reach of the spatial resolution of
existing measurement techniques.  
\end{abstract}

\pacs{72.80.Vp,68.37.Hk,05.30.Fk}


\maketitle

Nanometer resolution temperature measurements are technologically necessary, for instance, to characterize the
thermal performance and failure mechanisms of semiconductor devices \cite{Altet06}, or to investigate bioheat
transfer at the molecular level for the treatment of cancer or cardiovascular diseases \cite{Bischof06}.  Fundamentally, local
temperature measurements of quantum systems can elucidate the correspondence between phonon
\cite{Chen03,Ming10,Galperin07}, photon \cite{deWilde06,Yue11,Greffet07}, and electron temperature
\cite{Engquist81,Dubi09c,Bergfield13demon} measures.  Moreover, quantum effects 
may offer novel methods to
circumvent long-standing technological challenges, suggesting that the investigation of `phase sensitive'
\cite{Buttiker89} thermal effects could open the door to quantum engineered heat transport devices
\cite{Cahill03,Cahill02}.

%

Quantum coherent temperature oscillations have been predicted in 1-D ballistic systems \cite{Dubi09a,Dubi09b} and
in small conjugated organic molecules \cite{Bergfield13demon}, but despite impressive advances in thermal microscopy
\cite{Kim11,Yu11,Kim12,Fabian12} that have dramatically increased the spatial resolution of temperature
measurements, these predictions are not yet within reach of experimental verification.

In this letter, we investigate the local electron temperature distribution 
of carbon nanoribbon (CNR) and graphene
junctions covalently bonded to two metallic electrodes used to apply a thermal bias, and probed using a third scanning electrode
acting as a local thermometer.  
We find
that the Friedel oscillations and temperature oscillations in these systems are related, and that techniques to
modify the former \cite{Jiamin12} can also be used to modify the latter. Specifically, we investigate the response of
junctions to an applied gate voltage and find that the temperature oscillation wavelength can be varied over several
orders of magnitude, bringing these oscillations within the spatial resolution of current techniques in thermal microscopy \cite{Kim11,Yu11,Kim12,Fabian12}.


{\em Theory --}
Defining a local electronic temperature in a system out of equilibrium requires consideration of a local probe
(thermometer) that couples to the system and whose 
temperature is
varied until the local properties
of the system are minimally perturbed \cite{Dubi09a,Dubi09b,Dubi11}---a floating probe.
This should occur when the thermometer reaches {\em local equilibrium} with the system, i.e., when there is no longer
any net flow of charge or heat between the system and the probe \cite{Bergfield13demon}.
Several variations on the later condition have also been discussed in the literature \cite{Engquist81,Sanchez11,Jacquet11,Caso11}.
%
In terms of the
currents, the temperature of the probe is then defined by the conditions \cite{Bergfield13demon}
\begin{equation}
I_p^{(\nu)}=0, \;\; \nu=0,1,	
\label{eq:equil}
\end{equation}
where $-eI_p^{(0)}$ and $I_p^{(1)}$ are the charge and heat currents flowing into the probe, respectively. 

\begin{figure}[bt]
	\centering
		\includegraphics[width=2.5in]{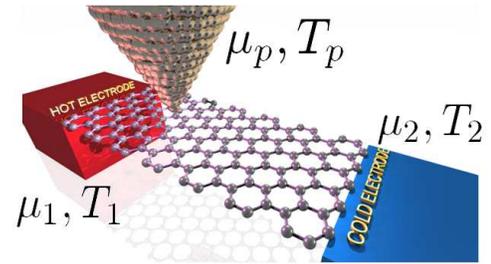}
	\caption{A schematic representation of a three terminal CNR junction with the hot and cold electrodes covalently bonded to the CNR and a third
scanning thermal probe positioned over the CNR.  The probe is allowed to come into thermal and electrical equilibrium with the sample and measure the
temperature $T_p$.}
	\label{fig:0129_schematic}
\end{figure}

We consider 
junctions composed of a CNR or graphene molecule, hot and cold electrodes
bonded to the molecule, a probe electrode, and the environment
(see Fig.\ \ref{fig:0129_schematic}).
The hot and cold electrodes provide a thermal gradient, but form an open electrical circuit in a thermal transport experiment.
Under these conditions, and in linear response, the heat current flowing into the scanning thermal probe is
\cite{Bergfield13demon}
\begin{equation}
I_p^{(1)}\!=\!
\sum_{\beta=1}^2 \tilde{\kappa}_{p\beta} (T_\beta \!-\! T_p)
+
\kappa_{p0}(T_0\!-\!T_p) + \kappa_{ph}(T_{ph} \! - \! T_p),
\label{eq:probe_I1}
\end{equation}
where $T_\beta$ is the temperature of terminal $\beta$, $\tilde{\kappa}_{\alpha\beta}$ is the 
thermal conductance between electrodes $\alpha$ and $\beta$, $\kappa_{p0}$ is the thermal
coupling of the probe to the ambient environment at temperature $T_0$, and $\kappa_{ph}$ is the phonon heat conductance between the probe and a phonon bath with temperature $T_{ph}$. 
The environment
could be, for example, the black-body radiation or
gaseous atmosphere surrounding the circuit, or the
cantilever/driver on which the temperature probe
is mounted \cite{Bergfield13demon}.

Eqs.\ (\ref{eq:equil}) and (\ref{eq:probe_I1}) can be solved for the temperature of a probe
in thermal and electrical equilibrium with, and coupled locally to the system of interest
\cite{Bergfield13demon}
\begin{equation}
T_p=\frac{\tilde{\kappa}_{p1} T_1 + \tilde{\kappa}_{p2} T_2 + \kappa_{p0} T_0 + \kappa_{ph}T_{ph}}{\tilde{\kappa}_{p1} +
\tilde{\kappa}_{p2}+  \kappa_{p0} + \kappa_{ph}}.
\label{eq:Tp}
\end{equation}
%
Here the thermal conductance $\tilde{\kappa}_{\alpha\beta}$ between electrodes $\alpha$ and $\beta$ within the three-terminal thermoelectric
circuit formed by the probe and hot and cold electrodes is \cite{Bergfield13demon}
\begin{align}
\tilde{\kappa}_{\alpha\beta} &= \frac{1}{T}\left[\Lfun{2}{\alpha\beta}
-\frac{\left[\Lfun{1}{\alpha\beta}\right]^2}{\tilde{\cal L}_{\alpha\beta}^{(0)}}
\right.
\nonumber \\
&- 
\left.
{\cal L}^{(0)} \!
\left(
\frac{\Lfun{1}{\alpha\gamma}\Lfun{1}{\alpha\beta}}{\Lfun{0}{\alpha\gamma}\Lfun{0}{\alpha\beta}}
+\frac{\Lfun{1}{\gamma\beta}\Lfun{1}{\alpha\beta}}{\Lfun{0}{\gamma\beta}\Lfun{0}{\alpha\beta}}
-\frac{
\Lfun{1}{\alpha\gamma}\Lfun{1}{\gamma\beta}
}{
\Lfun{0}{\alpha\gamma}\Lfun{0}{\gamma\beta}
}
\right)
\right],
\label{eq:kappatilde}
\end{align}
where $\Lfun{\nu}{\alpha\beta}$ is an Onsager linear-response coefficient,
$\tilde{\cal L}_{\alpha\beta}^{(0)}=\Lfun{0}{\alpha\beta}+
 \Lfun{0}{\alpha\gamma}\Lfun{0}{\gamma\beta}/(\Lfun{0}{\alpha\gamma}+\Lfun{0}{\gamma\beta})$
and $1/{\cal L}^{(0)} = 1/\Lfun{0}{12} + 1/\Lfun{0}{1p} + 1/\Lfun{0}{2p}$.

We envision experiments performed in ultrahigh vacuum (UHV) with the electronic temperature probe operating in the tunneling regime and scanned across the
sample at fixed height.
Under linear-response conditions,
electron-phonon interactions and inelastic scattering are weak in graphene, so the indirect
phonon contributions to $\Lfun{0}{\alpha\beta}$ and $\Lfun{1}{\alpha\beta}$
can be neglected.  
Thermal transport from phonons is included via $\kappa_{ph}$.
The linear response coefficients needed to evaluate Eq.\ (\ref{eq:Tp}) may thus be calculated using elastic electron transport theory
\cite{Sivan86,Bergfield09a}
\begin{equation}
\Lfun{\nu}{\alpha\beta}
= \frac{1}{h} \int dE \; (E-\mu_0)^{\nu}\,\myT_{\alpha\beta}(E) \left(-\frac{\partial f_0}{\partial E}\right),
\label{eq:Lnu}	
\end{equation}
where $f_0$ is the equilibrium Fermi-Dirac distribution with chemical potential $\mu_0$ and temperature $T_0$.
The transmission function \cite{
DiVentraBook,Bergfield09a} ${\myT}_{\alpha\beta}(E)={\rm Tr}\left\{
\Gamma^\alpha(E) G(E) \Gamma^\beta(E) G^\dagger(E)\right\}$ is expressed in terms of the 
tunneling-width matrices $\Gamma^\alpha$ and the retarded Green's function of the junction $G(E) = [{\bf S} E -
H_{\rm mol} - \Sigma_{\rm T}(E)]^{-1}$, where
the overlap matrix ${\bf S}$ reduces to the identity matrix in an orthonormal basis and
$\Sigma_{\rm T}(E) = -i \sum_\alpha \Gamma^\alpha(E)/2$.  
Throughout this work we consider transport in the wide-band limit where $\Gamma^\alpha(E) \approx
\Gamma^\alpha$.

%

\begin{figure*}[tb]
	\centering
	\includegraphics[width=6.5in]{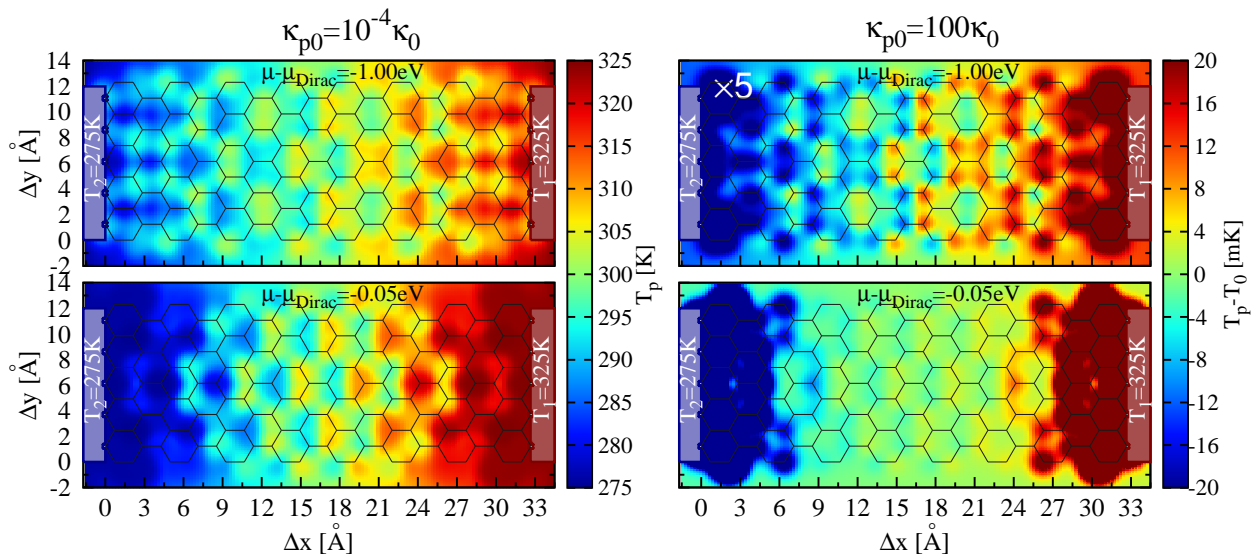}
		\caption{The calculated spatial temperature profile for an armchair CNR probed by a Pt SThM fixed 2.5\AA\ above the sheet shown for two energies and for weak and strong environmental coupling with $\kappa_{p0}$=$10^{-4}$$\kappa_0$ and $\kappa_{p0}$=100$\kappa_0$, respectively.  The values in the top right-hand panel are multiplied by a factor of five.  In all panels phonons are included with $\kappa_{ph}$=0.01$\kappa_0$.  By adjusting $|\mu - \mu_{\rm Dirac}|$ the temperature oscillation wavelength can be tuned.  Even with strong environmental coupling and significant phonon heat conductance, the quantum temperature oscillations are visible.  The phonon temperature $T_{ph}$ is taken to vary linearly between each electrode and the applied temperature gradient across the nanoribbon is 50K.}
	\label{fig:Armchair_CNR_temp}
\end{figure*}

In the vicinity of the Dirac point, a simple tight-binding Hamiltonian has been shown to accurately describe the
$\pi$-band dispersion of graphene \cite{Reich02}.  The molecular 
Hamiltonian is 
\begin{equation}
	H_{\rm mol} = \sum_{\langle ij \rangle} t_{ij} d_i^\dagger d_j + {\rm H.c},
\end{equation}
where $t=-2.7\mbox{eV}$ is the nearest-neighbor hopping matrix element between 2p$_z$ carbon orbitals of the graphene
lattice, and $d_i^\dagger$ creates an electron on the i$^{th}$ 2p$_z$ orbital.
To be specific, we consider here a scanning thermal microscope (SThM) with an atomically-sharp Pt tip
operating in the tunneling regime but near contact.
The tunneling-width matrix may be described in general as \cite{Bergfield12a}
$\Gamma^p_{nm} = 2\pi V_{n} V_m^\ast\, \rho_p$,
where $n$ and $m$ label $\pi$-orbitals of the molecule,
$\rho_p(E)$ is the local density of states on
the apex atom of the probe electrode, and $V_m$ is the tunneling matrix element 
between the quasi-atomic apex wavefunction and orbital $m$ of the molecule.  We consider all $s,p,d$ orbitals 
of the Pt SThM's apex atom and the $\pi$-system of the carbon sheet, meaning that the transport into the probe is multi-channel 
\cite{Bergfield12a}.

{\em Results --} The calculated local temperature distribution of an armchair CNR bonded to hot and cold electrodes held at
$T_1$=325K and $T_2$=275K, respectively, is shown for several gate potentials and environmental coupling strengths in Fig.\
\ref{fig:Armchair_CNR_temp}.  In these calculations, the SThM is scanned 2.5\AA\ above the plane of the carbon nuclei and the $\Gamma$ matrices describing the lead-molecule coupling are diagonal.  Non-zero elements of $\Gamma$, drawn as small red or blue circles in the figure, indicate contact between the electrode and the carbon atoms of the nanoribbon and are equal to 150meV.  
The probe is operating in the tunneling regime since the sum of Pt and C covalent radii is $\sim$2.03\AA\ \cite{CRC}. 
As indicated in the figure, the wavelength of the temperature variations changes as the quasiparticle energy is adjusted close to the Dirac point $\mu_{\rm Dirac}$.

In the simulations presented here, we consider
both a weak environmental coupling $\kappa_{p0}$=$10^{-4}\kappa_0$,
and 
a realistic environmental coupling
$\kappa_{p0}$=100$\kappa_0$, where $\kappa_0$=$(\pi^2/3)(k_B^2 T/h)$=0.284nW/K is the thermal conductance quantum at 300K
\cite{Rego98}.  
The weak coupling value $\kappa_{p0}$=$10^{-4}\kappa_0$ corresponds to the radiative coupling between a tip with effective radius $\sim$100nm and the blackbody environment, a fundamental limit on $\kappa_{p0}$ \cite{Bergfield13demon}.  At larger values of $\kappa_{p0}$, the amplitude of the quantum temperature oscillations is reduced due to the
reduced sensitivity of the thermal measurement \cite{Bergfield13demon,Kim12}, but the qualitative features of the interference pattern are preserved.
For comparison, the UHV SThM of Kim {\it et al.}\ \cite{Kim12} recently achieved $\kappa_{p0}$$\approx$700$\kappa_0$.  The phonon heat conductance $\kappa_{ph}$ is small since the Debye frequency of Pt and the CNR's phonon distribution are incommensurate and, at 2.5\AA\ above the CNR, the probe is not in contact with the CNR.  
We consider a realistic value of $\kappa_{ph}$=0.01$\kappa_0$, and let $T_{ph}$ vary linearly between the hot and cold electrodes.




\begin{figure}[tb]
\centering
\includegraphics[width=3.5in]{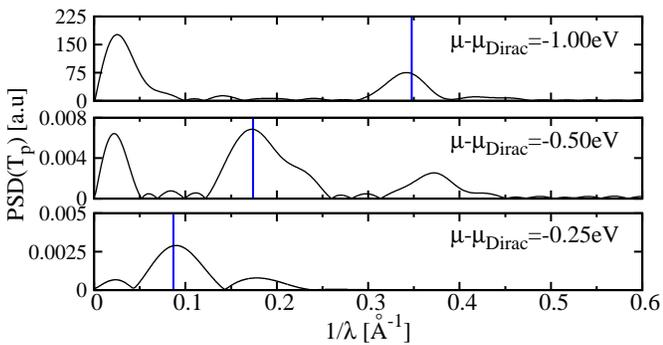}
\caption{The power spectral density (PSD) of a slice through the center row of the calculated temperature profiles shown in the left panels of Fig.\ \ref{fig:Armchair_CNR_temp}, with $\kappa_{p0} = 10^{-4}\kappa_0$.  The temperature oscillation wavelength increases as $\mu-\mu_{\rm Dirac}$ is decreased, in good agreement with Eq.\ \ref{eq:lambda}, whose values are indicated by vertical blue lines.  The PSD spectra are complex because of the small size of the CNR, the multi-mode nature of the Pt SThM, and the phonon conductance.  The temperature data within 3\AA\ of each electrode have been neglected in the PSD spectra. }
\label{fig:armchair_temperature_oscillations_vs_mu}
\end{figure}

The spatial temperature variations are a consequence of quantum interference~\cite{Dubi09c}, where the flow of heat from the hot and
cold electrodes into the probe is determined by position-dependent interferences and the molecular density of
states \cite{Bergfield13demon}.  According to Eq.\ (\ref{eq:Tp}), a maximally hot spot will be observed whenever
$\kappa_{p1}\gg \kappa_{p2}$, and vice versa for a maximally cold spot.  In general,
the largest variations in temperature will be observed when the thermal conductance from one of the two electrodes into
the probe is suppressed by destructive quantum interference \cite{Bergfield13demon}, which occurs when
the phase between thermal transport paths differs by $\pi$, so that $2k_F\Delta L = 2\pi$.
Such $2k_F$ oscillations are ubiquitous in electron systems at low temperatures, the best known example being the Friedel
oscillations in the density of states or charge density \cite{Jiamin12}.
%
%

\begin{figure}[tb]
	\centering
	\includegraphics[width=0.8\linewidth]{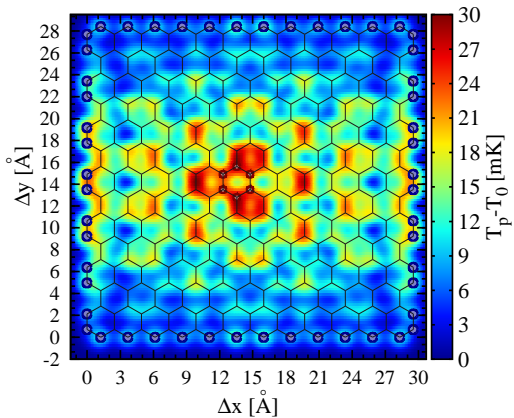}
\caption{The simulated temperature profile of a graphene fragment with a hot ($T_1$=350K) needle electrode 
(the benzene-like contact pattern is indicated with red dots) and a cold electrode ($T_2$=$T_0$=300K) bonded to the periphery of the sheet (blue dots) probed by a Pt SThM tip scanned 2.5\AA\ above the plane of the carbon nuclei.  In these simulations, we use $\kappa_{p0}$=700$\kappa_0$ (extracted from experiment) and $\kappa_{ph}$=0.01$\kappa_0$.  Here $\mu-\mu_{\rm Dirac}=-1\mbox{eV}$. 
The hot needle and periphery electrode have per orbital coupling strengths of 1eV and 0.1eV, respectively.  
}
	\label{fig:graphene_flake_temperature}
\end{figure}

Due to its unique dispersion relation,
the Friedel oscillation wavelength in graphene 
depends strongly on the energy of the
quasiparticles, which may be controlled via the application of a gate voltage \cite{Jiamin12}
\begin{equation}
	\lambda_{\rm Friedel}(E) = \frac{\hbar v_{\rm F}}{2 E},
	\label{eq:lambda}
\end{equation}
where $E$ is the energy away from the Dirac point.  In our tight-binding Hamiltonian $\hbar v_{\rm F}=3ta/2$, where
$t$=2.7eV is the tight-binding matrix element and $a$=1.42\AA\ is the C-C distance \cite{Neto09}.
The power spectral density (PSD) of a slice through the center row of 
the CNR shown in the left panels of Fig.\ \ref{fig:Armchair_CNR_temp} is shown for $\mu-\mu_{\rm Dirac}$=-1.00eV, -0.50eV, and -0.25eV in Fig.\ \ref{fig:armchair_temperature_oscillations_vs_mu}.  As shown in the figure, a spectral peak shifts as $\mu-\mu_{\rm Dirac}$ changes, in good agreement with Eq.\ \ref{eq:lambda} (shown as vertical blue lines in the figure).  
Closer to the Dirac point, where the Friedel oscillation wavelength becomes comparable to the linear dimensions of the system simulated, it is not
straightforward to resolve this peak above the background of peaks at small wavevectors arising from finite-size effects.  Nonetheless, it is clear from
Fig.\ \ref{fig:Armchair_CNR_temp} (lower panels) that the dominant wavelength of the temperature oscillations grows dramatically as 
$\mu \rightarrow \mu_{\rm Dirac}$.

The wide tunability of the temperature oscillations over orders of magnitude in wavelength in graphene
indicates that they are within the spatial resolution of current
SThM technology, which has achieved spatial and thermal resolution of 10nm and 15mK, respectively \cite{Kim12},
provided the phase coherence length of the carriers is sufficiently long.  
In pure graphene the dominant dephasing mechanism is deformation potential scattering by acoustic phonons
\cite{Hwang08}.  Using the scattering rate derived in Ref.~\onlinecite{Hwang08} and assuming that the momentum
relaxation time is equivalent to the phase-relaxation time, the phase-coherence length is given by
\begin{equation}
L_{\phi}(E) = \frac{4\hbar^3 \rho_m v_f^3 v_s^2}{D_{\rm A}^2 k_{\rm B} T E}
\end{equation}
where $D_A$ is the deformation potential, $v_{ph}$=2$\times10^{6}$cm/s is the acoustic phonon velocity,
$\rho_m\sim$7.6$\times10^{-8}$g/cm$^2$ is the graphene mass density, and
$v_f\sim$1.53$\times 10^{5}$m/s is the Fermi velocity.
The deformation potential reported in the literature typically ranges from 10-30eV.  As an example, with
$D_A$=30eV and $T$=300K, $L_{\phi}$(1.0eV)=68.4nm and $L_{\phi}$(0.05eV)=1.36$\mu m$.  These estimates, which are
in good agreement with 
recent experimental phase-coherence length measurements of carbon nanoribbons \cite{Minke12}, clearly indicate that quantum
thermal oscillations in graphene can occur on length scales well within the resolution of existing SThM techniques.
Indeed, the more formidable experimental challenge is likely to be reducing the environmental coupling $\kappa_{p0}$ of the probe
to increase the amplitude of the thermal oscillations above the threshold for observation (cf.\ Fig.\ \ref{fig:Armchair_CNR_temp}).

As a final example of an experimentally realistic system which may be used to investigate quantum temperature
oscillations, we consider a graphene flake with a hot needle-like terminal in the center, and the edge of the
flake held at ambient temperature. 
The temperature profile for this junction is
shown in Fig.\ \ref{fig:graphene_flake_temperature} for 
$\kappa_{p0}=700\kappa_0$, corresponding to the current experimental sensitivity \cite{Kim12}.
In Fig.\ \ref{fig:graphene_flake_temperature}, we have taken $\mu-\mu_{\rm Dirac}= -1\mbox{eV}$;  
the predicted temperature profile exhibits a strong dependence on gate voltage and exhibits quantum oscillations
within the resolution of current state-of-the-art SThM techniques.
We stress that although
computational resources have limited our discussion to small molecule structures, longer wavelength oscillations
should be observable in larger systems provided the transport is phase coherent and coupling to the environment
is minimized.


{\em Conclusion --}
We have found that temperature oscillations in carbon nanoribbon and graphene junctions, like
Friedel oscillations, can be tuned over orders of magnitude in wavelength,
making this an ideal system for both fundamental and device related studies into the nature of temperature and heat
transport at the nanoscale.

Work by J.P.B. and M.A.R. was supported as part of the Non-Equilibrium Energy Research Center (NERC), an Energy Frontier Research Center funded by the U.S. Department of Energy, Office of Science, Basic Energy Sciences under Award DE-SC0000989.  C.A.S. acknowledges support from the U.S. Department of Energy (DOE), Basic Energy Sciences 
under Award No. DE-SC0006699.  
M.D. acknowledges support from the DOE under Grant No. DE-FG02-05ER46204.


\bibliography{refs}

\end{document}